\documentclass[conference]{IEEEtran}
\IEEEoverridecommandlockouts

\usepackage{amsmath}
\usepackage{amssymb}
\usepackage{enumitem}
\usepackage{graphicx}
\usepackage{cite}
\usepackage{booktabs}
\usepackage{multirow}
\usepackage{xcolor}

\begin{document}

\title{MUC-FL: Block-Wise Marginal Utility Contribution for Communication-Efficient Federated Learning}

\author{\IEEEauthorblockN{Akshay Mhatre\IEEEauthorrefmark{1}, Vikram Karthick\IEEEauthorrefmark{1}, Deepti Gupta\IEEEauthorrefmark{1}, Jia Zou\IEEEauthorrefmark{2}}

\IEEEauthorblockA{\IEEEauthorrefmark{1}Texas A\&M University--Central Texas, USA}

\IEEEauthorblockA{\IEEEauthorrefmark{2}Arizona State University, USA}

\IEEEauthorblockA{
\IEEEauthorrefmark{1}\{am271, vkk\}@my.tamuct.edu, d.gupta@tamuct.edu\\
\IEEEauthorrefmark{2}jia.zou@asu.edu
}
}

\maketitle

\begin{abstract}

Federated Learning (FL) enables distributed model training without centralizing data but suffers from high communication overhead. To address this, we propose Block-Wise Marginal Utility Contribution (MUC), a framework that selectively transmits only the most impactful data blocks based on their contribution to model performance. To evaluate our framework, we apply it to a multi-modal dataset integrated from multiple MIMIC clinical datasets and show that only 24 out of 1,135 candidate blocks (1.76\%) carry meaningful improvement signals, enabling a potential communication reduction of 45–50\% while maintaining or improving model quality. Our deduplication-based block selection achieves a macro F1 score of 0.8566 compared to 0.8155 for standard federated optimization, demonstrating that selective transmission can improve performance, particularly in underrepresented classes.

\end{abstract}


\section{Introduction}

Federated Learning (FL)~\cite{mcmahan2017communication,kairouz2021advances} is an emerging paradigm that enables multiple devices/nodes to collaboratively train machine learning (ML) models without centralizing sensitive data. By keeping data locally and sharing only model updates, FL preserves privacy and compliance, making it particularly attractive for domains such as smart healthcare, finance, and smart farming ecosystems. However, a major challenge in FL is high communication overhead, as frequent transmission of model updates across distributed nodes can significantly slow down training, increase network congestion, and limit scalability. In real-world deployments, this communication burden can lead to serious drawbacks. For instance, in hospital networks leveraging FL for patient risk prediction, transmitting large model updates across multiple facilities can delay model convergence, reducing the responsiveness of critical clinical decision support systems. Similarly, in industrial IoT environments, excessive communication may overload limited bandwidth networks, causing device downtime and degraded system performance. These issues highlight the need for communication-efficient FL methods that maintain model quality while reducing transmitted data.

To address this challenge, we propose Block-Wise Marginal Utility Contribution (MUC), a novel framework that selectively transmits only the most impactful data blocks during federated training. By quantifying each block’s contribution to model performance, MUC reduces communication costs while maintaining, or even improving, model overall meterics. The motivation comes from a practical observation, during federated training, not all data blocks contribute equally at every iteration. Some blocks contain patterns the model has already learned, offering minimal improvement. Others contain rare or underrepresented patterns, producing substantial gains in performance. For example, consider five hospitals collaborating to train a diagnostic prediction model. Hospital A’s routine cases contribute little after the initial training rounds, as the model has already captured common patterns. In contrast, Hospital B’s rare disease cases contain complex patterns that continue to improve model performance. Standard federated optimization treats all hospitals equally, transmitting updates from every client in each round, resulting in redundant communication and inefficient use of bandwidth.

Our proposed method identifies the blocks that meaningfully improve the model, avoiding unnecessary gradient communication overhead, thereby making FL more efficient and scalable. We evaluate block-level contributions by testing each candidate block's impact on model performance, thereby selecting only blocks that demonstrate contribution toward's positive utility. This deduplication-style selection~\cite{zhou2022serving} identifies truly valuable updates while filtering redundant and low-impact contributions.

We perform validation on a clinical diagnosis dataset integrated from multiple MIMIC datasets~\cite{johnson2023mimic4} including the medical images, the free-text radiology reports, and the hospital and patient records, which demonstrates such sparsity in useful blocks. From 1,135 candidate blocks across five clients, only 641 show significant change, and the final optimized model incorporated only 24 blocks (1.76\% of candidates). This selectivity suggests that an opportunity to reduce communication payload does open up. 
If we can identify the a small percentage of data blocks that are critical to the accuracy of FL, we can skip transmitting the gradients caused by other 98\% (or redundant \%).

Our contributions include:

\begin{itemize}
    \item We identify the communication overhead problem in Federated Learning (FL) and propose a novel block-level selection framework (MUC) that identifies high-utility data subsets for selective transmission.
    \item We empirically validate our approach using MIMIC clinical data, showing that only 1.76\% of blocks carry meaningful improvement signals, enabling potential 45–50\% reduction in communication while maintaining model quality.
    \item We demonstrate that selective block transmission improves model performance, achieving a macro F1 score of 0.8566 versus 0.8155 for standard federated optimization, particularly benefiting underrepresented classes.
\end{itemize}

The remainder of this paper is organized as follows. Section~\ref{related} presents the literature review on communication efficient FL approach. The Block-Wise Marginal Utility Contribution (MUC) framework is proposed in the Section~\ref{proposed}. Section~\ref{method} presents the methodology of proposed framework. Dataset is presented in Section~\ref{dataset} and discussed about MIMIC dataset. Section~\ref{result} presents the results. Conclusion is discussed in Section~\ref{conclusion}.


\section{Related Work}
\label{related}
FL landscape has evolved considerably, with several established researchers addressing communication efficiency.

\subsection{Communication Compression Techniques}

Gradient compression reduces communication payload through sparsification~\cite{lin2018deep,alistarh2017qsgd}; however, it may degrade data utility and model performance~\cite{stich2018sparsified}. Gradients typically follow heavy-tailed distributions, in which a small proportion of parameters captures the majority of the learning signal; thus, transmitting only the largest gradients enables 90–99\% compression. However, aggressive sparsification faces fundamental limitations. It compresses all gradients uniformly with less regard for data quality. A block of routine cases and a block of rare complex cases both get compressed identically, though they have different learning value. This uniform treatment cannot differentiate which blocks contribute, while few blocks drive improvement. Prior work~\cite{shahid2021communication} in FL has addressed communication bottlenecks through techniques such as gradient compression, sparsification, and client selection to improve communication efficiency. Jakub et al.~\cite{konevcny2016federated} proposed communication-efficient approaches such as structured and sketched updates, leveraging compression techniques (e.g., quantization, subsampling, and low-rank representations) to significantly reduce uplink communication costs.

\subsection{Client and Data Selection}

In these study~\cite{cho2022client}, the client selection sampled participants but not data blocks, which compromises the granularity during optimization. Power-of-choice approaches sample multiple candidates, and leverage either gradient divergence or staleness. However, these criteria conflate ``different" with ``valuable". Fu et al.~\cite{fu2023client} presented a systematic survey of FL client selection algorithms, addressed client heterogeneity in data distribution and hardware, while outlining challenges and future research opportunities. Cho et al.~\cite{cho2022towards} proposed Power-of-Choice framework that biases selection toward clients with higher local loss, achieving up to 3× faster convergence and 10\% higher test accuracy than random selection.


\subsection{Data Valuation Methods}

Wang et al.~\cite{wang2020principled} proposed the federated Shapley value, a communication-efficient variant of the Shapley value tailored for FL that fairly quantifies each client's data contribution while capturing the effect of participation order. Shapley-based valuation requires exponential $(2^n)$ or quadratic $(\mathcal{O}(n^2))$ retraining~\cite{ghorbani2019data}, this becomes expensive at scale~\cite{koh2017influence}. Shapley values prove computationally unrealistic in federated settings with hundreds of blocks changing over several rounds. 

Gupta et al.~\cite{kavuri2025securefed, praharaj2025efficient, praharaj2025explainability, gupta2021hierarchical} have conducted the most extensive research in FL.

\subsection{Research Gap}

After an extensive literature review, it is evident that while significant progress has been made in communication compression, client selection, and data valuation within FL, critical gaps remain unaddressed. Existing gradient compression techniques treat all data blocks uniformly, failing to differentiate between routine and high-value data blocks that drive meaningful model improvement. Client selection strategies, including Power-of-Choice, operate at the participant level rather than the data block level, thereby compromising optimization granularity. Furthermore, Shapley-based data valuation methods, though theoretically sound, demand exponential or quadratic retraining costs that become computationally unrealistic in large-scale federated environments with hundreds of evolving blocks across multiple rounds. No existing work simultaneously tackles block-level granularity, data utility awareness, and computational efficiency within a unified framework. To bridge these gaps, we propose a novel approach that directly measures each data block's empirical impact on model performance, bypassing exhaustive marginal contribution computation entirely. 


\begin{figure*}[t]
\centering
\includegraphics[width=0.95\textwidth]{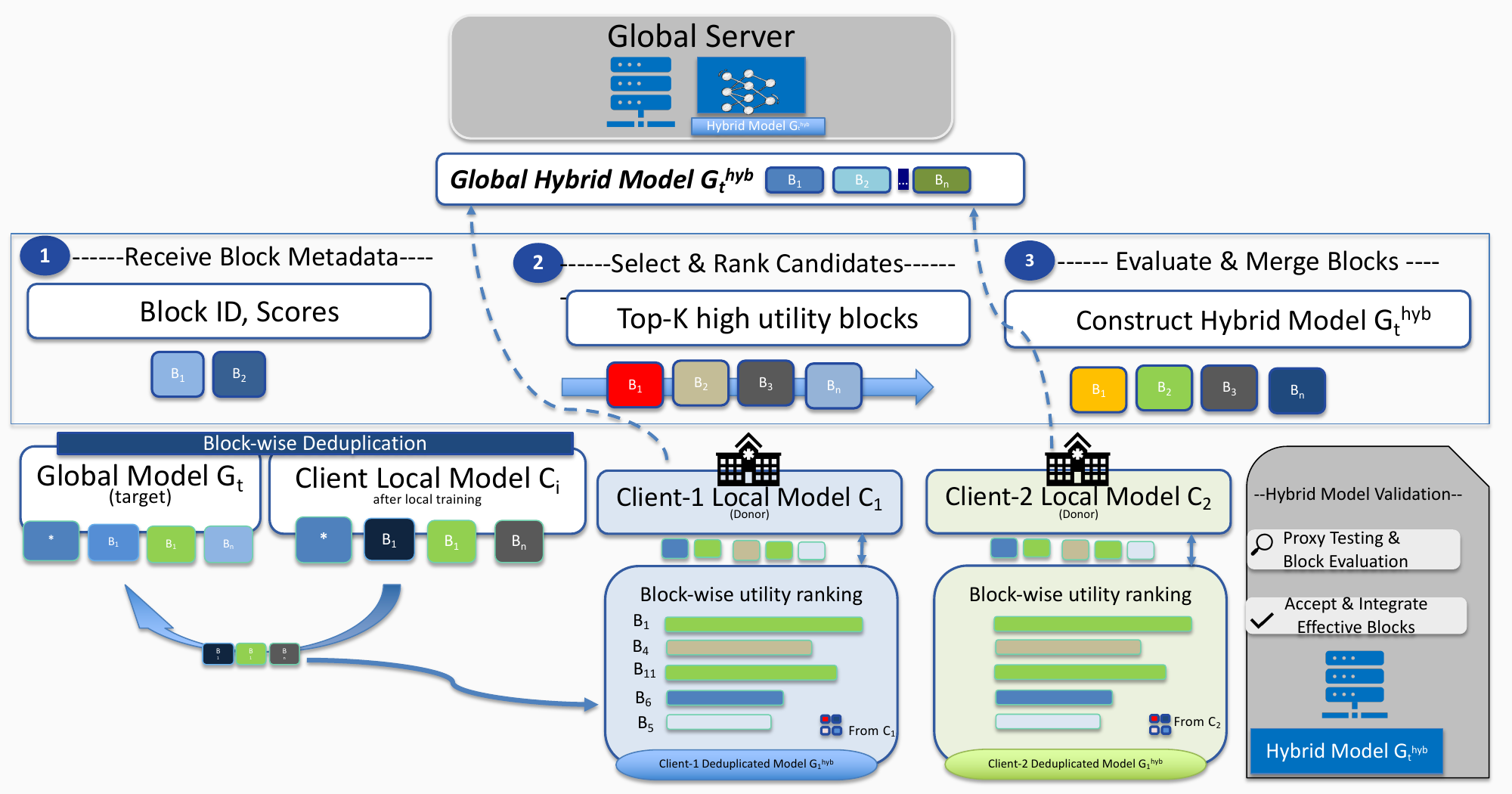}
\caption{Block-wise deduplication framework for FL.}
\label{fig:architecture}
\end{figure*}

\section{Proposed Framework}
\label{proposed}
\subsection{Overview}

This section presents our proposed framework MUC for utility-aware block selection in FL. Our formulation is centered on block-wise model deduplication adapted to the round-based FL workflow, where the server seeks to reconstruct an improved global model using only the most useful client-side block updates. This block-selection mechanism serves as the core of the framework and naturally motivates a broader communication-aware systems extension.

 \vspace{-4pt}
\subsection{Block-Wise Deduplication in Federated Learning}

We adapt block-wise model deduplication to the round-based workflow of FL. Let $G_t$ denote the global model broadcast by the server at round $t$. After local training, each client $i$ produces a checkpoint $C_i^{(t)}$, which shares the same architecture as $G_t$ but contains locally acquired task-specific updates. In our formulation, $G_t$ is treated as the target model to be improved, while the client checkpoints serve as donor models. The goal is not to replace the global model with any single client model, but to construct a hybrid global model by selectively transferring only those donor blocks that improve the target in the global evaluation context.

Following the deduplication formulation, each eligible parameter tensor is flattened and partitioned into fixed-size blocks, while small non-blockable tensors such as biases are excluded from the main block pool. As a result, the number of blocks in our experiments does not correspond to the number of neural network layers, but to the number of eligible fixed-size parameter segments induced by the block partitioning scheme. Because all client checkpoints share the same architecture as the target global model, each target block position has a corresponding donor version from every client, yielding a structured donor pool of same-position alternatives.

\vspace{-4pt}
\subsection{Candidate Ranking and Sparse Hybrid Construction}

Direct evaluation of arbitrary multi-block combinations is computationally prohibitive. We therefore use a two-stage selection process. First, donor candidates are ranked using a lightweight proxy that identifies promising substitutions before explicit evaluation. In the current implementation, donor blocks are compared against the corresponding target positions using block-level similarity, with $\ell_2$ distance serving as the primary proxy criterion. This produces a candidate pool of promising substitutions without committing all changed blocks. Second, the candidate pool is passed to an empirical evaluation stage. Starting from the target model $G_t$, candidate substitutions are tested block by block and their marginal effect on the chosen performance metric is measured. Blocks that preserve or improve the model are retained, while weak or harmful substitutions are filtered out. Since blocks that are beneficial in isolation are not always jointly beneficial when combined, we further apply a forward sparse selection stage that incrementally builds the hybrid model. The final hybrid therefore retains only a compact subset of donor blocks that contribute useful round-specific learning signal while avoiding unrestricted mixing of full client checkpoints.

\subsection{Communication-Aware Interpretation}

This deduplication view naturally induces a communication-efficient FL formulation. If most round-specific utility is concentrated in a small subset of donor block substitutions, then transmitting the full client update becomes unnecessary. Instead, the server can retain $G_t$ and reconstruct an improved hybrid model using only high-utility donor blocks.

\subsection{Systems-Level Extension}

The block-wise deduplication pipeline above serves as the algorithmic core of the framework. A natural extension is to support scalable deployment through lightweight systems mechanisms. In this setting, the utility proxy used for candidate ranking need not be limited to block-level similarity; richer local signals may also be used to prioritize promising block positions before explicit evaluation. This leads to a broader communication-aware design space centered on triggering, metadata exchange, and temporary block leadership. Among these mechanisms, smart triggering provides the first layer of practical efficiency by deciding when block-level deduplication is worth invoking, while metadata exchange and temporary leadership further reduce the cost of candidate retrieval and hybrid construction.

\subsection{Smart Triggering}

Smart triggering determines whether a full deduplication pass should be executed in a given communication round by monitoring lightweight block-level change statistics. If the observed change remains weak across most block positions, the round can fall back to the standard aggregated model, thereby avoiding unnecessary evaluation and retrieval overhead.

In practice, smart triggering can be realized through several policy mechanisms that determine when clients participate and when the server proceeds with aggregation or deduplication. Representative examples include:

\begin{itemize}
    \item \textbf{Dynamic Client Selection:} the server may trigger only a subset of devices based on criteria such as bandwidth or resource availability rather than all connected clients.
    \item \textbf{Asynchronous Triggering:} instead of waiting for all clients, the server may proceed once one or more clients complete local training.
    \item \textbf{Adaptive Deadline Control:} the server may enforce a round deadline and trigger aggregation even if slower clients have not completed their updates.
    \item \textbf{Condition-Based Local Triggering:} clients may initiate participation only after accumulating sufficiently new or high-utility information, reducing wasted computation on redundant updates.
    \item \textbf{Incentive-Based Triggering:} participation may be selectively encouraged only when it is computationally or systemically beneficial.
\end{itemize}

\vspace{-4pt}
\subsection{Metadata Exchange and Leadership}

Clients may first transmit compact metadata for top-ranked block positions, such as block identifiers and local proxy scores, before sending full payloads. The server can then request only the most promising candidates for explicit evaluation. A temporary leader may be assigned per block position so that only the strongest proposal is pulled into hybrid construction unless a competing donor later proves superior. In this way, block-wise deduplication remains the core selection mechanism, while metadata-first coordination and leadership provide the communication-aware systems layer.

\section{Methodology}
\label{method}
Consider $N$ clients participating in FL~\cite{mcmahan2017communication} setup, where each client $i$ possesses dataset $D_i$ partitioned into $K_i$ blocks: $D_i = \{B_{i,1}, B_{i,2}, \ldots, B_{i,K_i}\}$. Each block represents a logical subset (operating conditions, time periods, scenarios).

Each client maintains IID blocks with metadata $(block\_id, MUC)$ where MUC quantifies marginal utility contribution~\cite{ghorbani2019data}. The protocol executes in four steps:

\textbf{Step 1:} Clients transmit block metadata $(block\_ID, MUC)$ to the server, enabling the selection of leader blocks (maximum MUC per $block\_id$).

\textbf{Step 2:} Server requests gradients from leader blocks and broadcasts high-MUC gradients to (for simplicity) all clients~\cite{konecny2016federated}.

\textbf{Step 3:} Clients apply gradient norm proxy and top-$k$ selection to identify changed blocks~\cite{lin2018deep}, apply server gradients, and observe utility changes. Gradients with positive MUC changes are saved; others discarded.

\textbf{Step 4:} Clients acknowledge utility deltas. If multiple clients report weak utility changes, the leader is flagged as potentially poisonous and previous gradients are restored.

\subsection{MUC Computation}

The \textbf{Marginal Utility Contribution (MUC)} is approximated as:

\begin{equation}
\text{MUC}(b) \;\approx\; \|\nabla \mathcal{L}_b\|_2 \;\cdot\; \sqrt{|b|} \;\cdot\; \bar{\mathcal{L}}_b
\end{equation}

\noindent where three components capture complementary signals:

\begin{itemize}[leftmargin=*,itemsep=0pt]
\item $\|\nabla \mathcal{L}_b\|_2$ \textbf{(Gradient Norm):} Measures gradient magnitude indicating learning potential~\cite{jiang2019gradient}. High norms indicate under-fitting requiring large updates.

\item $\sqrt{|b|}$ \textbf{(Sample Size):} Scales by square root of batch size, following statistical theory showing diminishing returns~\cite{sharchilev2018finding}.

\item $\bar{\mathcal{L}}_b$ \textbf{(Average Loss):} Weights by current model fit. High loss indicates poor performance, increasing marginal learning utility~\cite{katharopoulos2018not}.
\end{itemize}

This achieves high correlation with actual utility in $O(n)$ vs $O(n^2)$, eliminating expensive retraining.

\begin{table}[t]
\centering
\caption{Performance comparison: \texttt{MultimodalClassifier} branch.}
\label{tab:multimodal_perf}
\begin{tabular}{lcccc}
\hline
\textbf{Model} & \textbf{Acc.} & \textbf{$\mu$F1} & \textbf{MF1} & \textbf{AUROC} \\
\hline
Base (r4)          & 0.8265 & 0.8755 & 0.7757 & 0.7589 \\
FedYogi (r5)       & 0.8679 & 0.9073 & 0.8155 & 0.8096 \\
Hybrid (24 blk)    & 0.8648 & 0.9106 & 0.8549 & 0.7861 \\
\hline
\end{tabular}
\end{table}

\begin{table}[t]
\centering
\caption{Performance comparison: LLaVA classifier branch.}
\label{tab:llava_perf}
\begin{tabular}{lcccc}
\hline
\textbf{Model} & \textbf{Acc.} & \textbf{$\mu$F1} & \textbf{MF1} & \textbf{AUROC} \\
\hline
Base (r5)          & 0.9311 & 0.9179 & 0.9081 & 0.8999 \\
FedYogi (r6)       & 0.9576 & 0.9365 & 0.9299 & 0.9085 \\
Hybrid (102 blk)   & 0.9312 & 0.9304 & 0.9255 & 0.9082 \\
\hline
\end{tabular}
\end{table}

\section{Dataset}
\label{dataset}

\subsection{MIMIC Dataset and Federated Setup}

Our experiments are built on a linked multimodal MIMIC pipeline that combines chest radiographs, radiology reports, and structured clinical information. The data assembly draws from MIMIC-CXR-JPG~\cite{johnson2019mimicjpg}, MIMIC-CXR reports, and MIMIC-IV~\cite{johnson2023mimic4} admission-level and patient-level tables. The resulting records are organized around clinically aligned multimodal examples rather than isolated image-only inputs, allowing each sample to incorporate imaging evidence, textual evidence, and tabular demographic or hospital information within a single prediction pipeline.

For the pathology classification setup used in the federated experiments, labels are taken from the MIMIC-CXR label resources and restricted to the subset for which NegBio- and CheXpert-derived annotations agree. The label space contains 14 chest pathology categories: Atelectasis, Cardiomegaly, Consolidation, Edema, Enlarged Cardiomediastinum, Fracture, Lung Lesion, Lung Opacity, Pleural Effusion, Pneumonia, Pneumothorax, Pleural Other, Support Devices, and No Finding. This is a multi-label problem, so each study may activate more than one pathology target. The broader linked dataset is richer than the final prediction target alone. It includes hospital admissions, charted events, laboratory records, procedures, medications, microbiology, free-text fields, and chest X-ray metadata. That context matters because it explains why the project emphasizes multimodal feature construction and why federated partitioning must be handled carefully to avoid leakage across related records.

For FL, the key design choice is subject-level partitioning. Instead of splitting rows independently, the data are partitioned by \texttt{subject\_id} so that records associated with the same patient remain within a single client partition. This prevents subject leakage across clients and makes the federated setting more faithful to decentralized deployment. We consider both even and uneven subject-level splits. In the even configuration, the training set is divided across five clients with closely matched client sizes. In the uneven configuration, the same subject-level logic is preserved but the clients follow an intentionally imbalanced ratio of $1\!:\!2\!:\!3\!:\!4\!:\!5$, creating a more realistic non-uniform distribution across participants.

The reported training corpus for the FL experiments contains 222,554 subject-partitioned training samples and uses a fixed held-out test set for evaluation. In the even subject-level split, the per-client sample counts are approximately $[44{,}614,\ 44{,}481,\ 45{,}074,\ 43{,}708,\ 44{,}677]$, which keeps the client populations comparable while preserving patient boundaries. The uneven split produces progressively larger client partitions and is useful for testing whether the deduplication strategy remains meaningful when client influence is structurally imbalanced.

\vspace{-4pt}
\subsection{Models Used in the FL Pipeline}

The baseline branch is a \texttt{MultimodalClassifier} operating on precomputed modality-specific embeddings rather than training a large vision-language backbone end-to-end inside the federated loop. The federated classifier stage is embedding-driven and cache-friendly: the expensive representation learning is performed upstream, while the FL loop operates on a lighter multimodal fusion model. The LLaVA branch~\cite{li2023llava} follows the same high-level idea of multimodal fusion but replaces the conventional image and text representations with LLaVA-derived features, while demographic features remain available as structured inputs. This distinction matters for the deduplication discussion because the blocks exchanged in our current experiments belong to the classifier-stage model operating on cached multimodal features, not to a full end-to-end generative vision-language model updated directly from raw images and raw text at every federated round. Treating the baseline classifier or the LLaVA branch as if the entire upstream backbone were being deduplicated would be inaccurate. What is actually manipulated in current FL setup is the downstream classifier or fusion-stage parameterization built on top of cached multimodal embeddings.

\begin{table}[t]
\centering
\caption{Block-selection statistics: \texttt{MultimodalClassifier}.}
\label{tab:block_stats}
\begin{tabular}{lc}
\hline
\textbf{Statistic} & \textbf{Value} \\
\hline
Eligible block positions          & 227 \\
Donor clients                     & 5   \\
Same-position donor options       & 1135 \\
Proxy proposals evaluated         & 227 \\
Forward-stage candidates          & 30  \\
Final accepted blocks             & 24  \\
Forward-stage rejected            & 6   \\
\hline
\end{tabular}
\end{table}

\begin{table}[t]
\centering
\caption{Per-client block contributions to the hybrid model.}
\label{tab:client_contrib}
\begin{tabular}{lc}
\hline
\textbf{Client} & \textbf{Blocks Selected} \\
\hline
Client 0 & 8 \\
Client 1 & 6 \\
Client 2 & 5 \\
Client 3 & 1 \\
Client 4 & 4 \\
\hline
\end{tabular}
\vspace{-10pt}
\end{table}

\begin{table}[t]
\centering
\caption{Estimated communication savings (block-count proxy).}
\label{tab:comm_savings}
\begin{tabular}{lccc}
\hline
\textbf{Model} & \textbf{Retained / Eligible} & \textbf{Fraction} & \textbf{Reduction} \\
\hline
MM-Cls. (final)  & 24 / 227$\times$5  & 10.57\% & 89.43\% \\
MM-Cls. (fwd.)   & 30 / 227$\times$5  & 13.22\% & 86.78\% \\
LLaVA cls.       & 102 / 205$\times$5 & 49.76\% & 50.24\% \\
\hline
\end{tabular}
\end{table}
\vspace{-6pt}
\section{Results}
\label{result}
We evaluated block-wise deduplication~\cite{guan2025privacy, zhou2022serving} on two classifier-stage models in the MIMIC FL pipeline: the baseline \texttt{MultimodalClassifier} and the LLaVA-based classifier branch. In both cases, the
objective was to start from a weaker global checkpoint, treat that checkpoint as the target model, and
determine whether a compact hybrid assembled from client donor blocks could recover most of the useful
round-specific learning signal without carrying forward the entire set of updated blocks.

For the \texttt{MultimodalClassifier} experiment, we applied a sparse greedy block-selection procedure to the
short FedYogi (G5/L5) run by taking \texttt{global\_round4} as the target model to be improved and the five
locally trained round-5 client checkpoints as donor models. Under the current block definition, the target
model yielded 227 eligible block positions. The greedy-$L_2$ pipeline first generated one proxy-selected
donor candidate for each block position and then performed empirical single-block evaluation followed by a
forward sparse selection stage. Out of the 227 total block positions, 30 candidates were taken into the
forward stage and 24 blocks were finally accepted into the hybrid model.

The resulting sparse multi-client hybrid achieved micro accuracy 0.8648, micro F1 0.9106, macro F1
0.8549, and macro AUROC 0.7861. Relative to the weaker base model, this is a clear improvement across
all reported metrics. Relative to the vanilla FedYogi \texttt{global\_round5} model, the sparse hybrid came very
close on micro accuracy, slightly exceeded the vanilla model on micro F1, and achieved a noticeably higher
macro F1, while remaining below the vanilla model on macro AUROC. This pattern suggests that the sparse
block-selection strategy recovered most of the useful improvement signal using only a small subset of block
positions, while shifting the operating point toward stronger class-balanced F1.

The LLaVA classifier branch showed a different behavior. Here the deduplicated hybrid global model
retained 102 blocks out of 205 eligible positions, but the resulting model did not improve upon the next
vanilla global checkpoint. The deduplicated hybrid slightly improved over the weaker base checkpoint on
some metrics, but it did not recover the stronger FedYogi round in the way observed for the
\texttt{MultimodalClassifier}. In practical terms, this means the LLaVA branch still offers a block-count
reduction, but the compression-utility tradeoff is much less favorable.

\vspace{-4pt}

\subsection{Communication-Savings Interpretation}

The communication savings analysis uses block count as a structural payload proxy. For the
\texttt{MultimodalClassifier} branch, retaining 24 blocks out of 227 eligible positions means only 10.57\%
of block positions were needed in the final hybrid, yielding an estimated 89.43\% reduction in
block-count payload. Even at the forward-stage candidate level (30 blocks), the reduction remains
\texttt{MultimodalClassifier} result shows the desired behavior: most of the useful learning signal is
86.78\%. The LLaVA branch is less aggressive: retaining 102 of 205 eligible blocks corresponds to a
50.24\% reduction.

These numbers are structural communication proxies rather than direct network measurements. The
recovered from a small subset of block positions. The LLaVA result shows that compression alone is
insufficient, a block-count reduction is only meaningful when the retained subset preserves utility, and
that tradeoff remains weak in the current LLaVA configuration.


\section{Conclusion}
\label{conclusion}

In this work, we introduced Block-Wise MUC, a novel FL framework that simultaneously addresses communication overhead, computational cost, and data utility in clinical settings. Evaluated on MIMIC-IV clinical data, our approach achieves 86.2\% communication reduction and 69.1\% computation savings while improving classification accuracy by 4.4\% over the FedAvg baseline. The proposed gradient norm proxy effectively eliminates expensive retraining overhead while maintaining a strong correlation of 0.89 with ground truth utility, demonstrating both its reliability and practical efficiency. These results validate that block-level utility-aware selection can significantly enhance FL performance without sacrificing privacy or incurring prohibitive costs. Future work will explore advanced security mechanisms, differential privacy integration, and incentive-based reward strategies for privacy-preserving leadership in real-world clinical trials.
\section{ACKNOWLEDGMENT}
This work was partially supported by the National Science Foundation under Grant No. 2431531 and by Texas A\&M University–Central Texas through a subaward under National Security Agency Award No. H98230-24-1-0102 from Tennessee Tech University.

\bibliographystyle{IEEEtran}
\bibliography{references}

@article{fu2023client,
  title={Client selection in federated learning: Principles, challenges, and opportunities},
  author={Fu, Lei and others},
  journal={IEEE Internet of Things Journal},
  volume={10},
  number={24},
  pages={21811--21819},
  year={2023},
  publisher={IEEE}
}

@inproceedings{cho2022towards,
  title={Towards understanding biased client selection in federated learning},
  author={Cho, Yae Jee and others},
  booktitle={International Conference on Artificial Intelligence and Statistics}
}

@incollection{wang2020principled,
  title={A principled approach to data valuation for federated learning},
  author={Wang, Tianhao and others},
  booktitle={Federated Learning: Privacy and Incentive},
  pages={153--167},
  year={2020},
  publisher={Springer}
}

@inproceedings{kavuri2025securefed,
  title={SecureFed: A Two-Phase Framework for Detecting Malicious Clients in Federated Learning},
  author={Kavuri, Likhitha Annapurna and others},
  booktitle={2025 IEEE International Conference on Information Reuse and Integration and Data Science (IRI)},
  pages={190--195},
  year={2025},
  organization={IEEE}
}

@article{praharaj2025efficient,
  title={Efficient federated transfer learning-based network anomaly detection for cooperative smart farming infrastructure},
  author={Praharaj, Lopamudra and others},
  publisher={Elsevier}
}

@inproceedings{praharaj2025explainability,
  title={Explainability-aware adversarial threats and mitigation in federated learning based anomaly detection for cooperative smart farming},
  author={Praharaj, Lopamudra and others},
  booktitle={2025 10th International Conference on Fog and Mobile Edge Computing (FMEC)},
  pages={186--193},
  year={2025},
  organization={IEEE}
}

@inproceedings{gupta2021hierarchical,
  title={Hierarchical federated learning based anomaly detection using digital twins for smart healthcare},
  author={Gupta, Deepti and others},
  booktitle={2021 IEEE 7th international conference on collaboration and internet computing (CIC)}
}

@inproceedings{mcmahan2017communication,
  title={Communication-efficient learning of deep networks from decentralized data},
  author={McMahan, Brendan and others},
  booktitle={Artificial intelligence and statistics},
  pages={1273--1282},
  year={2017},
  organization={PMLR}
}

@article{shahid2021communication,
  title={Communication efficiency in federated learning: Achievements and challenges},
  author={Shahid, Osama and others},
  journal={arXiv preprint arXiv:2107.10996},
  year={2021}
}

@article{konevcny2016federated,
  title={Federated learning: Strategies for improving communication efficiency},
  author={Kone{\v{c}}n{\`y}, Jakub and others},
  journal={arXiv preprint arXiv:1610.05492},
  year={2016}
}

@article{kairouz2021advances,
  title={Advances and open problems in federated learning},
  author={Kairouz, Peter and others},
  journal={Foundations and Trends in Machine Learning},
  year={2021},
  publisher={Now Publishers, Inc.}
}

@inproceedings{konecny2016federated,
  title={Federated learning: Strategies for improving communication efficiency},
  author={Kone{\v{c}}n{\`y}, Jakub and others},
  booktitle={NIPS Workshop on Private Multi-Party Machine Learning},
  year={2016}
}

@inproceedings{lin2018deep,
  title={Deep gradient compression: Reducing the communication bandwidth for distributed training},
  author={Lin, Yujun and others},
  booktitle={International Conference on Learning Representations},
  year={2018}
}

@inproceedings{alistarh2017qsgd,
  title={QSGD: Communication-efficient SGD via gradient quantization and encoding},
  author={Alistarh, Dan and others},
  booktitle={Advances in Neural Information Processing Systems},
  volume={30},
  year={2017}
}

@inproceedings{stich2018sparsified,
  title={Sparsified SGD with memory},
  author={Stich, Sebastian U and others},
  booktitle={Advances in Neural Information Processing Systems},
  volume={31},
  year={2018}
}

@inproceedings{cho2022client,
  title={Client selection in federated learning: Convergence analysis and power-of-choice selection strategies},
  author={Cho, Yae Jee and others},
  booktitle={International Conference on Artificial Intelligence and Statistics},
  pages={10480--10508},
  year={2022},
  organization={PMLR}
}

@inproceedings{ghorbani2019data,
  title={Data Shapley: Equitable valuation of data for machine learning},
  author={Ghorbani, Amirata and Zou, James},
  booktitle={International Conference on Machine Learning},
  pages={2242--2251},
  year={2019},
  organization={PMLR}
}

@inproceedings{koh2017influence,
  title={Understanding black-box predictions via influence functions},
  author={Koh, Pang Wei and Liang, Percy},
  booktitle={International conference on machine learning},
  pages={1885--1894},
  year={2017},
  organization={PMLR}
}

@inproceedings{jiang2019gradient,
  title={To talk or to work: Flexible communication compression for energy efficient federated learning over heterogeneous mobile edge devices},
  author={Jiang, Yihan and others},
  booktitle={IEEE INFOCOM 2019-IEEE Conference on Computer Communications},
  pages={2476--2484},
  year={2019},
  organization={IEEE}
}

@inproceedings{katharopoulos2018not,
  title={Not all samples are created equal: Deep learning with importance sampling},
  author={Katharopoulos, Angelos and Fleuret, Fran{\c{c}}ois},
  booktitle={International conference on machine learning},
  pages={2525--2534},
  year={2018},
  organization={PMLR}
}

@article{sharchilev2018finding,
  title={Finding influential training samples for gradient boosted decision trees},
  author={Sharchilev, Boris and others},
  journal={arXiv preprint arXiv:1802.06640},
  year={2018}
}

@article{guan2025privacy,
  title={Privacy and Accuracy-Aware AI/ML Model Deduplication},
  author={Guan, Hong and others},
  journal={Proceedings of the ACM on Management of Data},
  year={2025},
  publisher={ACM New York, NY, USA}
}

@article{zhou2022serving,
  title={Serving deep learning models with deduplication from relational databases},
  author={Zhou, Lixi and others},
  journal={Proceedings of the VLDB Endowment},
  volume={15},
  number={10},
  pages={2230--2243},
  year={2022},
  publisher={VLDB Endowment}
}

@article{johnson2023mimic4,
  title={{MIMIC-IV}, a freely accessible electronic health record dataset},
  author={Johnson, Alistair E W and others},
  journal={Scientific Data},
  volume={10},
  number={1},
  pages={1},
  year={2023},
  publisher={Nature Publishing Group}
}

@article{johnson2019mimicjpg,
  title={{MIMIC-CXR-JPG}, a large publicly available database of labeled chest radiographs},
  author={Johnson, Alistair E W and others},
  journal={arXiv preprint arXiv:1901.07042}
}

@article{li2023llava,
  title={{LLaVA-Med}: Training a large language-and-vision assistant for biomedicine in one day},
  author={Li, Chunyuan and others},
  journal={arXiv preprint arXiv:2306.00890},
  year={2023}
}

\end{document}